# Remote Energetic Neutral Atom Imaging of Electric Potential Over a Lunar Magnetic Anomaly

Y. Futaana[1]*, S. Barabash[1], M. Wieser[1], C. Lue[1,2], P. Wurz[3], A. Vorburger[3], A. Bhardwaj[4], K. Asamura[5]

[1]Swedish Institute of Space Physics, Box 812, SE-98128 Kiruna, Sweden.

[2]Department of Physics, Umeå University, Linnaeus väg 24, SE-90187 Umeå, Sweden.

[3]Physikalisches Institut, University of Bern, Sidlerstrasse 5, CH-3012 Bern, Switzerland.

[4]Space Physics Laboratory, Vikram Sarabhai Space Center, Trivandrum 695 022, India.

[5]Institute of Space and Astronautical Science, 3-1-1 Yoshinodai, Sagamihara 252-5210, Japan.

*Corresponding author: Yoshifumi Futaana (futaana@irf.se)

**Abstract**.

The formation of electric potential over lunar magnetized regions is essential for understanding fundamental lunar science, for understanding the lunar environment, and for planning human exploration on the Moon. A large positive electric potential was predicted and detected from single point measurements. Here, we demonstrate a remote imaging technique of electric potential mapping at the lunar surface, making use of a new concept involving hydrogen neutral atoms derived from solar wind. We apply the technique to a lunar magnetized region using an existing dataset of the neutral atom energy spectrometer SARA/CENA on Chandrayaan-1. Electrostatic potential larger than +135 V inside the Gerasimovic anomaly is confirmed. This structure is found spreading all over the magnetized region. The widely spread electric potential can influence the local plasma and dust environment near the magnetic anomaly.

## 1. Introduction

The electrostatic potential between the Moon surface and space is a key parameter that is fundamental for lunar science and human exploration on the Moon. Many investigations of electric potential and associated electric fields have been conducted from the surface using solar wind plasma since the Apollo era. Surface potential influences ambient plasma characteristics and electric current, which balances each other to produce an equilibrium state (Whipple, 1981). A large electric potential, sometimes less than 1 kV, was found from electron energy spectra when high energy electrons precipitated onto the lunar surface (Hakelas et al., 2005). The resulting induced current may potentially damage lander components. The surface potential also affects the environment by influencing dust dynamics and vice versa (e.g. Nitter et al., 1998; Stubbs et al., 2011; Garrick-Bethell et al., 2011).

On the other hand, a positive potential is not generally very large. For example, Freeman et al. [1973] used in situ measurement of suprathermal ions and found approximately +10V surface potential. Goldstein [1974] used electron data to find the -3 to +5V potential. These results are consistent with theoretical predictions of a few to +20V (e.g. Freeman and Ibrahim, 1975). The exception has been found inside the local magnetized regions (magnetic anomalies) on the lunar surface (e.g. Barnes et al., 1971; Hood et al., 1979; Tsunakawa et al., 2010), where a positive electric potential of the order of ~100 V is expected (Saito et al., 2012). Such electric potentials modify the local plasma environment significantly near the anomaly. Evaluating the electrostatic environment near magnetic anomalies is important as a possible candidate of landing site for future lander missions where cosmic ray protections are expected.

44    Energetic neutral atoms (ENAs), neutral atoms with energies more than 10 eV, have been
45 used for diagnostics for plasma and neutral environments. The powerful remote sensing
46 technique has provided the diagnostics with spatial scales from planetary bodies (Futaana et al.,
47 2011) to the Solar System (McComas et al., 2012). In the case of the Moon, the lack of an
48 intrinsic magnetic field (Colburn et al., 1967; Dyal et al., 1974) allows the solar wind to interact
49 directly with the surface, where ENAs are produced. Thus, ENAs provide information about the
50 solar wind access at the lunar surface (Futaana et al., 2006). The lunar ENAs have already been
51 detected by the ENA sensor, CENA (Chandrayaan-1 Energetic Neutral Atoms), on board a lunar
52 orbiter, Chandrayaan-1 (Barabash et al., 2009). These ENAs, composed only of hydrogen, are
53 originally solar wind protons, which are neutralized on and backscattered from the lunar surface
54 (Wieser et al., 2009).

55    The physical mechanism of the backscattered ENA generation at the lunar surface is not
56 fully understood. The biggest unknown is what causes the high backscattering fraction of 10–
57 20% of impinging solar wind proton flux (Wieser et al., 2009; McComas et al., 2009; Rodríguez
58 et al., 2012; Futaana et al., 2012). The observations contradict with classical understandings of
59 full (<0.1%) absorption (e.g. Behrisch and Wittmaack, 1991; Feldman et al., 2000; Crider and
60 Vondrak, 2002) because of the high porosity of the surface. Another unexplained signature is the
61 ENA energy spectrum. The observed energy spectrum of the backscattered ENAs follows the
62 Maxwell-Boltzmann distribution function with the characteristic energy, i.e. the temperature of
63 the distribution, ranging from 60 to 140 eV. A Maxwell-Boltzmann distribution implies a state of
64 thermal equilibrium, but it is unrealistic to assume equilibrium with such a high temperature (100
65 eV corresponds to $1.16 \times 10^6$ K) at the surface. Moreover, the characteristic energy depends only
66 on the solar wind energy; actually, the observation exhibits a linear correlation with the solar

67  wind velocity (Futaana et al., 2012). From classical theory of elastic scattering (e.g. Niehus et al.,
68  1993) linearity with the incident energy (i.e. the solar wind energy) is expected. From these
69  perspectives, an adaption of laboratory knowledge and theoretical understandings of the
70  backscattered process to the situation of a regolith surface in space is needed.

71  Even though the observed characteristics of the backscattered ENAs are not yet fully
72  explained by present theory, we propose here a new empirical method to derive the lunar surface
73  potential from observed hydrogen ENA energy spectra. In this paper, we first describe a new
74  method to map the surface electrostatic potential. Then, we apply this method to available
75  observations near a magnetic anomaly to show a large potential generation over a wide range of
76  the magnetic anomaly. We also discuss the result and its influence on surface plasma and dust
77  environments and human activities.

78  **2. Mapping method of electric potential**
79  Assuming the existence of a surface potential ($\pm\Phi_{surf}$) between the lunar surface and
80  space (solar wind), the solar wind will be decelerated (or accelerated) before reaching the
81  surface. The solar wind energy at the surface becomes $E_{surf}=E_{sw}-(\pm\Phi_{surf})$, where $E_{sw}$ is the
82  original solar wind energy. When the decelerated (or accelerated) solar wind is backscattered as
83  ENAs, the generated ENA temperature, $T_{ena}$, is determined only from the solar wind energy at
84  the lunar surface, $E_{surf}$ (Futaana et al., 2012). Because $T_{ena}$ is a measurable quantity, we can
85  determine $E_{surf}$. The difference between $E_{surf}$ and $E_{sw}$ provides the surface potential, $\pm\Phi_{surf}$.

86  For this method, it is important to produce a proper reference model, i.e., the relationship
87  between $E_{sw}$ and $T_{ena}$ for uncharged surface. We here use a model derived from more than 100
88  CENA observations in the equatorial region (Futaana et al., 2012):

89    $Tena = Vsw \times 0.273 - 1.99$

90   where *Tena* is mearued in eV and *Vsw* is the solar wind velocity in km/s. This can be rewritten

91   as:

92   $$Esw = \frac{(Tena + 1.99)^2}{14.19} \quad (eq.1)$$

93   where *Esw* is also measured in eV. This reference model is derived empirically. No theoretical

94   assurance is yet provided. Here we estimated a statistical uncertainty using the dataset in

95   Futaana et al. [2012] and calculated the ambiguity of the current reference model (eq1) to be -45

96   and +35 eV (as 25% and 75% percentiles, respectively). Note that the reference model is made

97   using the nominal dayside measurements near the equator. Because no strong magnetic

98   anomalies exist near the equator, the reference model is not affected by magnetic anomalies. On

99   the other hand, the reference model does not account for the surface potential in the nominal

100  dayside conditions. In the nominal dayside, the ultraviolet and electron illumination on the

101  surface produces a potential of a few eV (Vondrak et al., 1983; Němeček et al., 2011). We

102  assume in the following analysis that the surface potential in the nominal state is negligible

103  compared with that in the magnetic anomaly. Our results show that this assumption is realistic.

104        This new method relies on only the energy spectrum of ENAs. Therefore, it complements

105  classical methods using energy spectra of charged particles. Advantageously, the new method

106  provides an electrostatic potential map. Obviously, when a proper reference model is found, this

107  method can be applied to other airless bodies as Mercury, asteroids and Galileo moons.

108  **3. Application to magnetized region**

109        We applied the above described electric potential mapping method to the region over a

110  lunar magnetic anomaly called Gerasimovic (Fig. 1A). While there are many magnetized regions

111  on the Moon (Tsunakawa et al., 2010), the Gerasimovic is an isolated anomaly, and therefore, it

112  is suitable for a dedicated analysis and discussion (Vorburger et al., 2012). We used the data
113  obtained from the CENA sensor on the Chandrayaan-1 spacecraft over 8 non-consecutive orbits
114  from 16 June 22:06 to 18 June 00:28, 2009. The Moon was located in the solar wind. From
115  WIND/SWE observations, the solar wind is stable (density is ~6 cm$^{-3}$ and the velocity is ~300
116  km s$^{-1}$).

117     As seen in the ENA flux map (Fig. 1B), there are three characteristic regions (Wieser et
118  al., 2010) from the ENA flux (a) outside the anomaly, (b) enhanced region and (c) inside the
119  anomaly. Outside the anomaly, the ENA flux depends on, to the first order, the cosine of the
120  solar zenith angle, as the solar wind flux at the surface does. It corresponds to the drops in ENA
121  flux at high latitudes (Fig. 1B). For the electrostatic potential map (Fig. 1A), a clear signature
122  with a positive potential of >150 V is found inside the anomaly. The potential structure spread
123  over the majority of the magnetized area. In contrast, in the enhanced region and outside the
124  anomaly, no potential structures were found. Small-scale potential structures (below -100 or
125  above 100 V) are most likely artificial, because the spatial scale of them is comparable to the
126  dimension of the sensor FOV projection (~1° in latitude and ~8° in longitude).

127     Outside the anomaly, the characteristic energy of backscattered ENA is 84.3eV (Fig. 2).
128  Using the reference model (eq.1), the corresponding solar wind energy is found to be 525 eV. In
129  the enhanced region, the ENA flux is higher than in the other regions, and the ENA temperature
130  inside the enhanced region does not change (83.3 eV), indicating that the solar wind energy at
131  the surface of the enhanced region is 512 eV. Inside the magnetic anomaly, the ENA
132  characteristic energy is 72.5 eV, and the derived solar wind energy at the surface is 390 eV. The
133  difference in the solar wind energy between the outside (525 eV) and inside of the anomaly (390

134	eV) is 135 eV, which corresponds to the deceleration of the solar wind by electrostatic potential
135	above the magnetic anomaly.

136	The feature of above energy spectra, namely lower characteristic energy above
137	magnetized region, is commonly seen in the CENA dataset. Therefore, this empirical method of
138	electric potential derivation could also work for other magnetic anomalies, in which the surfaces
139	are positively charged.

140	**4. Discussion**

141	The positive potential formed above a magnetic anomaly is expected based on the charge
142	separation theory. Ions can penetrate farther than electrons in the interaction region, and the
143	charge separation produces an outward-facing electric field. Recently, Saito et al. [2012]
144	demonstrated that the *in situ* measurements of the energy spectra for solar wind protons, alpha
145	particles and electrons at 30 km altitude are consistently explained if one assumes the presence
146	of a +150 V electric potential inside the magnetic anomaly. These authors analyzed a different
147	anomaly (South Pole Aitken), but their electric potential is in a good agreement with our result of
148	135 V electric potential formation.

149	We emphasize here that we do not see any strong electric potential in the enhanced
150	region of the magnetic anomaly. The enhanced region can be attributed to an increase in the solar
151	wind proton flux at the lunar surface caused the deflection of the ion flow above the magnetic
152	anomaly (Wieser et al., 2010). The deflection magnifies the net solar wind flux of the enhanced
153	region, similar to the Earth's magnetosheath. However, the lack of electrostatic potential in the
154	enhanced region indicates that there is no electric potential formed above the enhanced region.
155	Therefore, the deflection above the anomaly is mainly caused by magnetic forces, and the

156　previously suggested mini-scale bow shock (Lin et al., 1998) is evidently not formed above the
157　magnetic anomaly.

158　　　　The large electric potential influences the near-surface environment near the magnetic
159　anomaly. It explains the observed signatures of solar wind ion reflection that are correlated with
160　the magnetic anomaly (Lue et al., 2011). A high electric potential not only decelerates the
161　incoming solar wind protons but also thermalizes the plasma and partially reflects the protons
162　(Saito et al., 2012) as illustrated in Fig. 3. This also modifies the lunar dust environment because
163　charged dust particles are lifted and transported by this large electric potential. Recently, the
164　differentiation of dust by a large electric potential was proposed (Garrick-Bethell et al., 2011;
165　Wang et al., 2012) to explain the coincidence of the magnetic anomalies and the coinciding
166　characteristic albedo signatures (swirls). Our finding of a large positive electric potential
167　structure all over the Gerasimovic anomaly supports the hypothesis of coinciding swirl
168　formation.

169　　　　Despite its effects on the near-surface environment, this relatively large electrostatic
170　potential above the magnetic anomaly does not pose any significant challenges for human and
171　robotic activities on the Moon. The widely spread electric potential structure over the magnetized
172　region suggests a relatively weak electric field. For example, assuming a 150V electrostatic
173　potential along 200 km, the corresponding electric field is 0.8 mV/m, which is of the same order
174　of the solar wind convection electric field at 1 AU. The vertical field could be stronger: If we
175　assume 10 km, the electric field could be 15 mV/m. This is still not strong enough to influence
176　the human and robotic activities directly, for example discharging at the lunar surface. Therefore,
177　a region under a magnetic anomaly still can be a good candidate site for landing and future
178　exploration of the Moon. Secondary effects, such as dust levitation due to the electric field and

179  its adsorption to any components (Stubbs et al., 2007), should be carefully assessed for robotic

180  activities, but such effects are commonly observed everywhere on the Moon.

181  **Acknowledgments**

182      This work was supported by funding from the Swedish National Space Board (SNSB) in

183  Sweden. We thank the WIND/SWE team for providing the solar wind data.

184

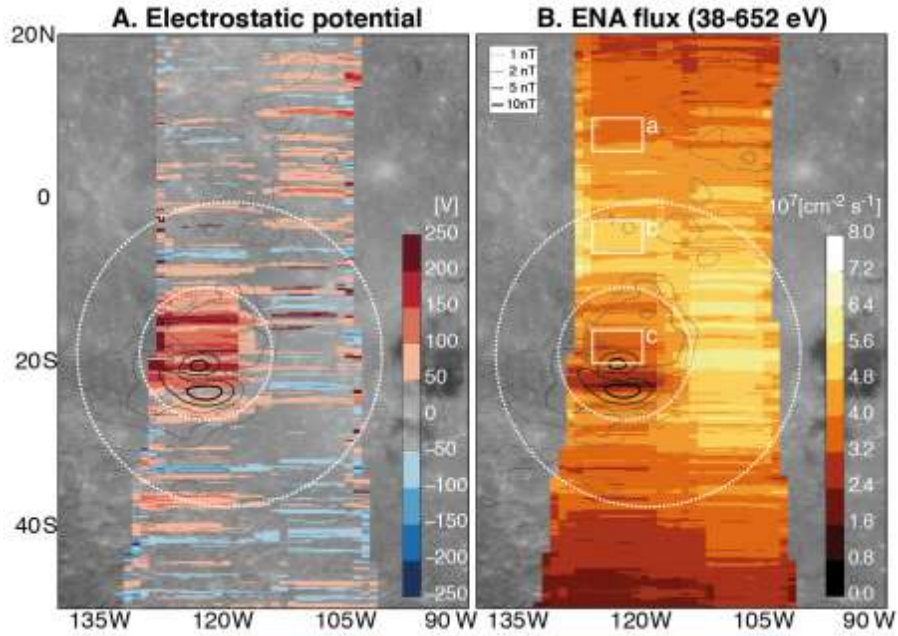

**Fig. 1**. **(A)** Map of the electrostatic potential near the Gerasimovic magnetic anomaly. Color scale shows the electrostatic potential with respect to the solar wind. Two dashed circles separate the regions inside magnetic anomaly, the enhanced region and the region outside the anomaly. **(B)** Map of the ENA flux integrated over 38–652 eV. Similar to the signatures in the integral flux over 150–600 eV reported previously (Wieser et al., 2010), three regions can be distinctly identified. Labeled white boxes indicate the areas that produce the energy spectra shown in Fig. 2. The Moon images are from the Clementine grayscale albedo map (Archinal et al., 2005). The contour lines represent the strength of the modeled magnetic field of anomalies at 30 km altitude (Purucker and Nicholas, 2010).

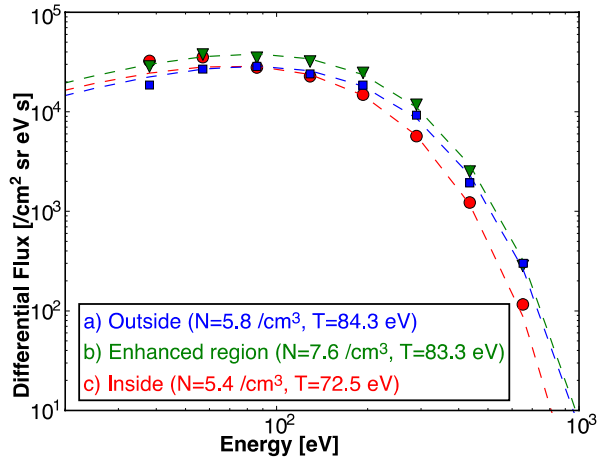

277

278   **Fig. 2 (Figure corrected).** Energy spectra of the backscattered ENAs from three characteristic

279   regions. The blue, green and red lines correspond to the backscattered ENAs from a) outside the

280   anomaly, b) the enhanced region and c) inside the anomaly, respectively (Fig. 1B). The symbols

281   show the flux observed by the CENA sensor for each region, and the dashed lines illustrate the

282   best fit by the Maxwell-Boltzmann distributions.

283

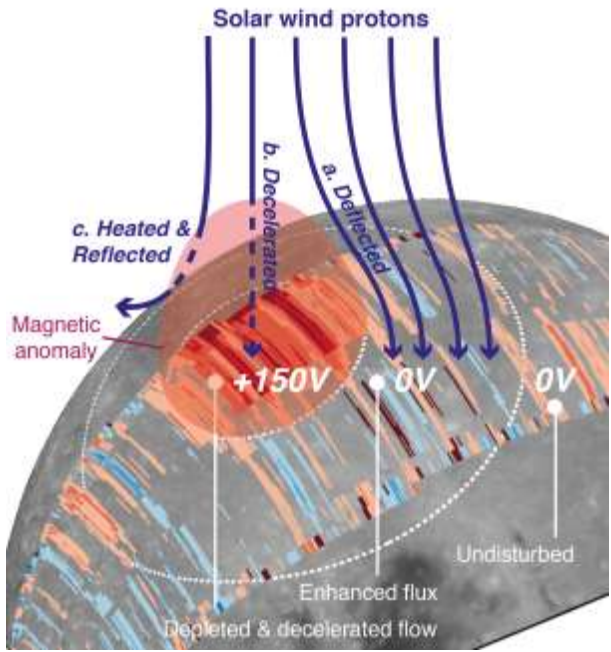

284

285　**Fig. 3.** An illustration of the solar wind interaction with the lunar magnetic anomaly. Blue lines

286　are solar wind proton streamlines modified by the interaction with the magnetic anomaly. The

287　consequences of the incoming solar wind protons are: a) deflected and reached the enhanced

288　region without a change in velocity, b) decelerated inside the magnetic anomaly due to the

289　potential structure of +150V and reached inside magnetic anomaly, or c) heated and reflected in

290　space before reaching the lunar surface.

291

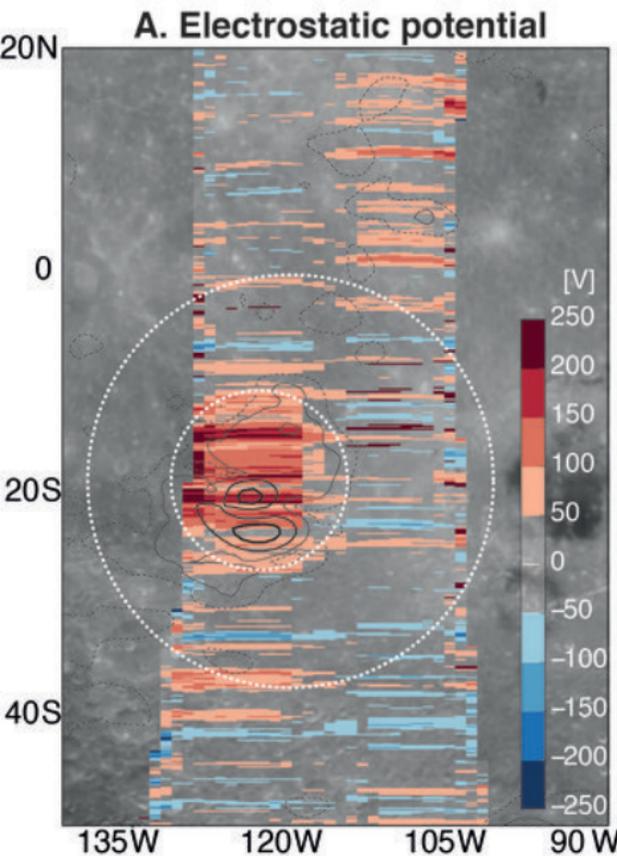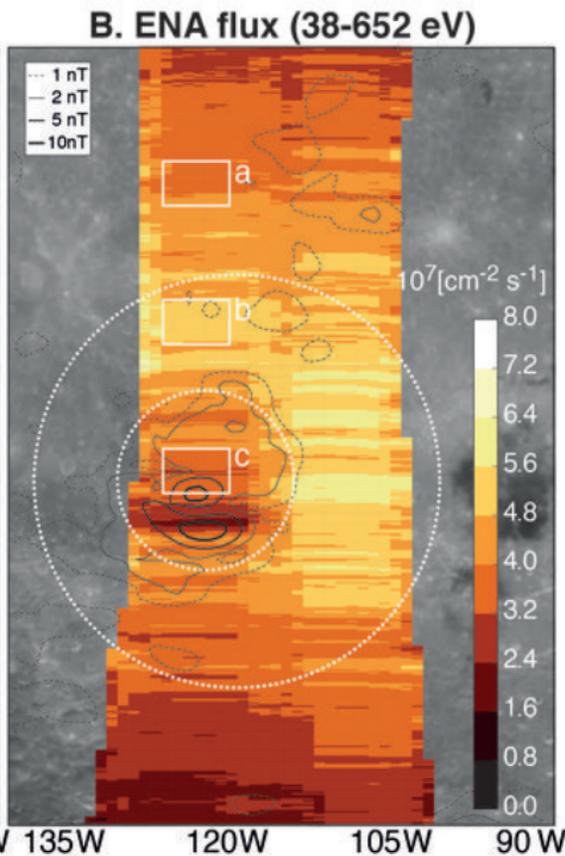

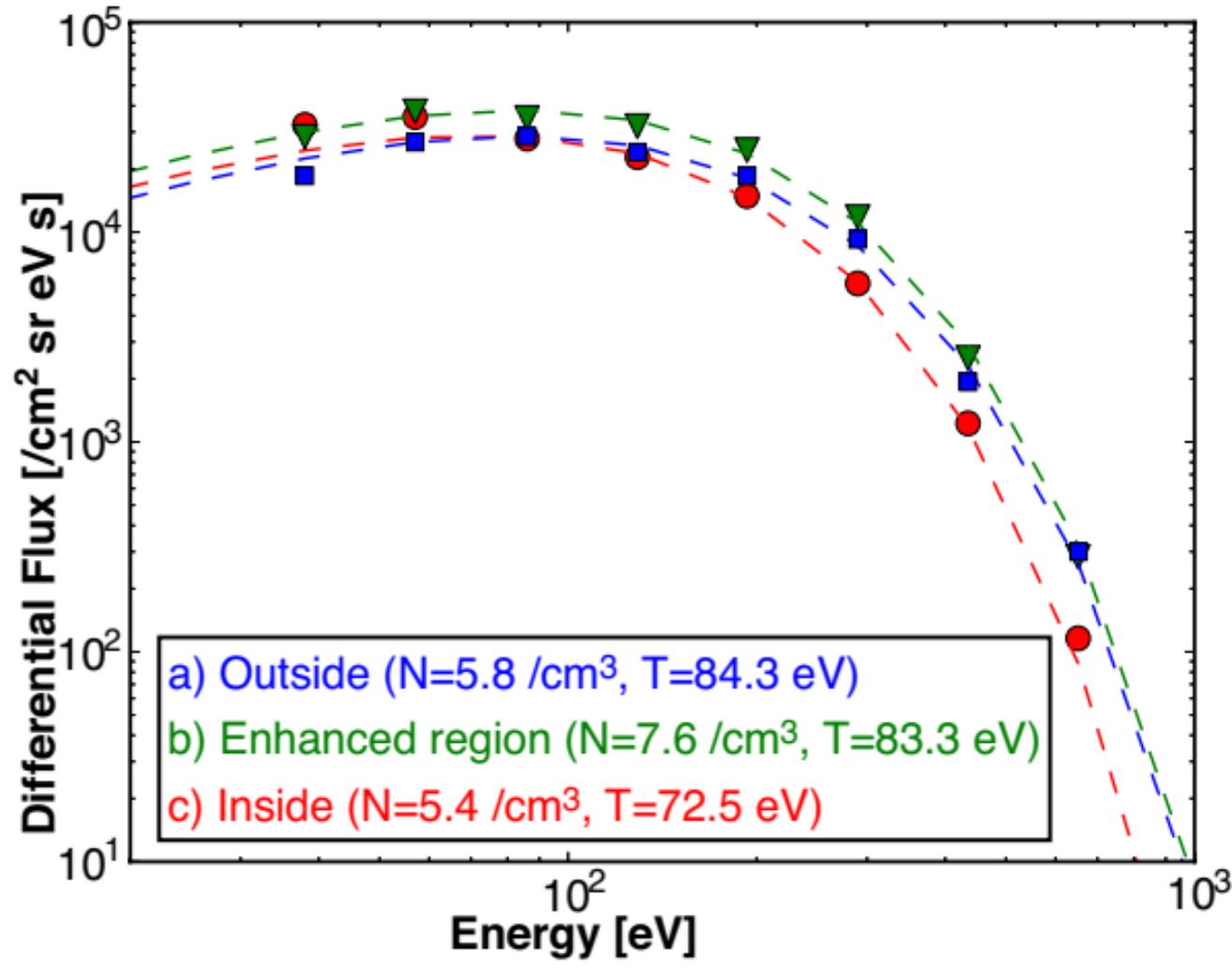

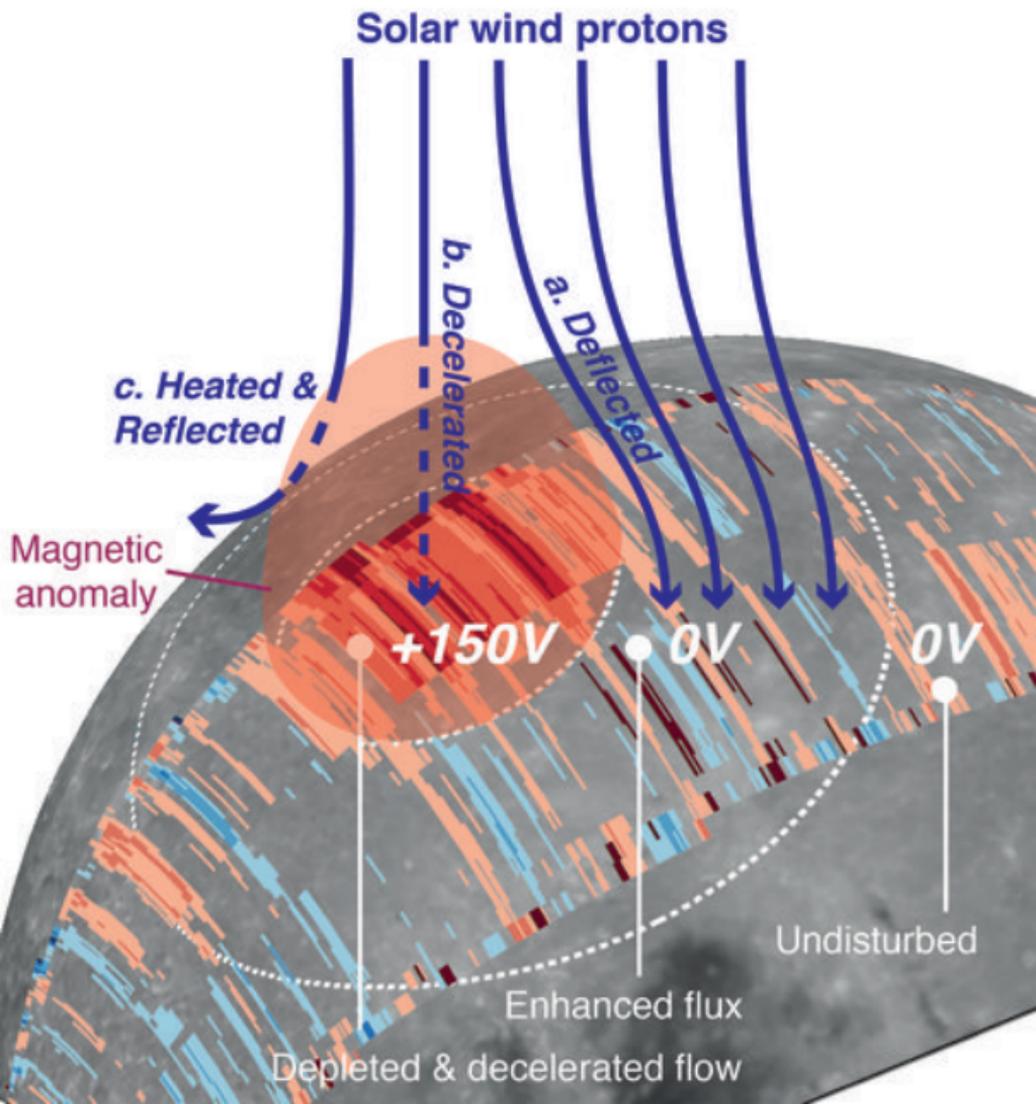